\documentclass[12pt,preprint]{emulateapj}

\slugcomment{For submission to ApJ}

\shorttitle{Cosmic Energy Spectrum}
\shortauthors{Driver et al.}

\begin{document}

%% LaTeX will automatically break titles if they run longer than
%% one line. However, you may use \\ to force a line break if
%% you desire.

\title{The energy output of the Universe from $0.1$ $\mu$m to $1000$ $\mu$m}

%% Use \author, \affil, and the \and command to format
%% author and affiliation information.
%% Note that \email has replaced the old \authoremail command
%% from AASTeX v4.0. You can use \email to mark an email address
%% anywhere in the paper, not just in the front matter.
%% As in the title, use \\ to force line breaks.

\author{Simon~P.~Driver\altaffilmark{1},
Cristina~C.~Popescu\altaffilmark{2}, Richard~J.~Tuffs\altaffilmark{3},
Alister~W.~Graham\altaffilmark{4}, Jochen~Liske\altaffilmark{5},
Ivan~Baldry\altaffilmark{6}}

\altaffiltext{1}{Scottish Universities' Physics Alliance (SUPA), School of Physics and Astronomy, 
University of St Andrews, North Haugh, St Andrews, Fife, KY16 9SS, UK; spd3@st-and.ac.uk}
\altaffiltext{2}{Centre for Astrophysics, University of Central Lancashire, 
Preston, PR1 2HE, UK}
\altaffiltext{3}{Max-Planck-Institut f\"ur Kernphysik, Saupfercheckweg 1, 69117 
Heidelberg, Germany}
\altaffiltext{4}{Centre for Astrophysics and Supercomputing, Swinburne University
of Technology, Hawthorn, Victoria 3122, Australia}
\altaffiltext{5}{European Southern Observatory, Karl-Schwarzschild-Str.~2, 85748
Garching, Germany}
\altaffiltext{6}{Astrophysics Research Institute, Liverpool John Moores
  University, Twelve Quays House, Egerton Wharf, Birkenhead CH4 1LD, UK}

%% Notice that each of these authors has alternate affiliations, which
%% are identified by the \altaffilmark after each name.  Specify alternate
%% affiliation information with \altaffiltext, with one command per each
%% affiliation.

%% Mark off your abstract in the ``abstract'' environment. In the manuscript
%% style, abstract will output a Received/Accepted line after the
%% title and affiliation information. No date will appear since the author
%% does not have this information. The dates will be filled in by the
%% editorial office after submission.

\begin{abstract}
The dominant source of electromagnetic energy in the Universe today
(over ultraviolet, optical and near-infrared wavelengths) is
starlight.  However, quantifying the amount of starlight produced has
proven difficult due to interstellar dust grains which attenuate some
unknown fraction of the light. Combining a recently calibrated
galactic dust model with observations of 10,000 nearby galaxies we
find that (integrated over all galaxy types and orientations) only
$(11 \pm 2)$\% of the $0.1$ $\mu$m photons escape their host galaxies;
this value rises linearly (with $\log \lambda$) to $(87 \pm 3)$\% at
$2.1$ $\mu$m. We deduce that the energy output from stars in the
nearby Universe is $(1.6\pm0.2) \times 10^{35}$ W Mpc$^{-3}$ of which
$(0.9\pm0.1) \times 10^{35}$ W Mpc$^{-3}$ escapes directly into the
inter-galactic medium. Some further ramifications of dust attenuation
are discussed, and equations that correct individual galaxy flux
measurements for its effect are provided.
\end{abstract}


\section{Introduction} The cosmic spectral energy distribution (CSED; 
e.g., Primack, Bullock \& Sommerville 2005) provides a description of
the current total (electromagnetic) energy output of the Universe over
all wavelengths. In the ultraviolet, optical and near-infrared
wavebands the CSED is dominated by starlight and its measurement can
be used to constrain the current stellar mass density and cosmic
star-formation rate as well as models of galaxy formation
\citep[e.g.,][]{baldry03,hopkins06}. The CSED is measured by
constructing the galaxy luminosity function (GLF)
\citep{schechter76,felten77,mgc05} at a specified wavelength (or
bandpass). The first moment of the GLF (extrapolated to bright and
faint magnitudes), gives the total luminosity density at this
wavelength and hence provides a single datum on the CSED.

Constructing the full CSED therefore requires accurate measurements of
the GLF in a variety of bandpasses. However, galaxies contain dust,
which, while negligible in terms of mass \citep{mgc07}, attenuates some
unknown fraction of the starlight before it exits a galaxy into the
inter-galactic medium (IGM) \citep{seares31,giovanelli95}. The
severity of this effect has proven difficult to quantify
\citep{disney89,valentijn90,burstein91}, leading to large
uncertainties in individual galaxy flux measurements and consequently
a systematic underestimation of the luminosity density (or individual
CSED measurements). The degree to which the CSED is underestimated
will, of course, be wavelength dependent \citep{cardelli89,calzetti01}
and it will depend critically on the amount and distribution of the
interstellar dust grains within the host galaxy (and an individual
galaxy's orientation to our line-of-sight).

Since the heated debate of the 90s, direct evidence from a number of
methods have led to the perspective that galaxies are predominantly
optically thin (at least in their outer regions). In particular,
evidence from overlapping galaxies \citep{white00}, have strongly
supported the stance that galaxies are optically thin, at least in the
inter-arm regions \citep{holwerda07}. Further detailed modeling of
extensive optical data on edge-on galaxies also appeared to support
the view that galaxies were optically thin throughout
\citep{xilouris99}. However, the optically thin case has proven
difficult to reconcile with the observed high level of far-infrared
emission which is presumed to arise from dust reradiating the
attenuated starlight \citep{bianchi00,popescu00,misiriotis01}. The
resolution to this conflict may lie in a more complex dust
distribution whereby galaxies may contain both optically thick (core
and arm) regions and optically thin (inter-arm and outer)
regions. \citet{tuffs04} and \citet{popescu00} (hereafter TP) have
advocated a three component dust model (optically thick inner disc,
thin outer disc and clumpy components), which is capable of
reproducing the detailed multiwavelength surface photometry from UV to
far-IR of edge-on galaxies such as NGC891 and the other galaxies
modelled by \cite{xilouris99}.

Recently, \citet{mgc07} identified evidence for strong and
inclination-dependent attenuation in a large sample of discs and
bulges which is not anticipated (or reproduceable) in models with a
purely optically thin dust distribution (e.g., see fig~5 in Popescu \&
Tuffs 2007). Similar and related results, albeit lacking bulge-disc
decompositions, have also now been reported for SDSS data \citep[e.g.,
Choi, Park \& Vogeley 2007;][]{shao07, unterborn08,
maller08,padilla08}. Using the TP-model we were able to reproduce the
attenuation-inclination relation for both discs and bulges and to
constrain the mean {\it central} $B$-band face-on opacity (the only
free parameter) for the galaxy population at large:
$\tau^{f}_{B}=3.8\pm0.7$. In this Letter we explore the implications
of this result on estimates of the CSED and ask whether the high value
for the central opacity can be reconciled with the far-IR output from
the nearby galaxy population. Throughout we adopt a standard flat
cosmology with $\Omega_{\rm M} = 0.3$, $\Omega_{\Lambda} = 0.7$ and
$H_0 = 70$~$h_{70}$~km~s$^{-1}$~Mpc$^{-1}$.

\section{Dust and its impact on galaxy photometry}
The dust model we adopt here is described fully in a sequence of
papers \citep{popescu00,tuffs04,mollenhoff06} and is summarised by
\citet{popescu07}. In brief the model incorporates three distinct
components: an extended optically thin dust disc associated with the
neutral hydrogen and older stellar population (i.e., the outer disc
and inter-arm regions); a less extended optically thick dust layer in
the spiral arms associated with the molecular hydrogen and younger
stellar population (i.e., primarily the inner disc); and a clumpy
component (representing star-forming molecular clouds).  Ideally
the optically thick component should be distributed according to a
spiral pattern, to better mimick the observed variation in opacity
between arm and inter-arm regions. For the purposes of this work this
distinction, between an additional disc of uniform opacity or one with
a built in spiral density pattern is not particularly relevant as both
imply a uniform high opacity disc in the central bulge-dominated
regions.

In \citet{mgc07} we constrained the TP-model's single free parameter,
the central face on opacity in $B$, which was found to be
$\tau^f_B=3.8\pm0.7$. Briefly, this constraint was obtained by
measuring the luminosity functions of galaxy bulges and discs at
various inclinations and comparing the dependence of the turn-over
point on inclination (i.e., the $M^*$--$\cos(i)$ relation) to
predictions of the TP-model (see \citealp{mgc07} for full details).

Having constrained the TP-model we can now determine the
inclination-dependent attenuation correction at any wavelength. From
Fig.~1 we see that the attenuation for bulge starlight can be as high
as $2$~B~mag (i.e., only $6$\% of the photons escape), and as high as
$1.2$~B~mag for discs (i.e., only $33$\% of the photons escape),
depending on wavelength and inclination. While the level of
attenuation of the disc is relatively consistent with previous
estimates, the bulge attenuation is a surprise and has not previously
been considered in detail. This is because while bulges are
traditionally considered to be dust free, like elliptical galaxies,
the dust in the galaxy disc attenuates bulge light, particularly from
the far-side. We can parameterize the dust attenuation curves of
Fig.~1 as follows:
\begin{eqnarray}
M^c_{\rm bulge} & = & M^o_{\rm bulge} - b_1 - b_2 [1-\cos(i)]^{b_3} \\ 
M^c_{\rm disc} & =& M^o_{\rm disc} - d_1 - d_2 [1-\cos(i)]^{d_3} 
\end{eqnarray} 
where $M^o$ and $M^c$ represent the observed and corrected magnitudes,
respectively, $i$ is the disc inclination, and the coefficients $b_1$,
$b_2$, $b_3$ and $d_1$, $d_2$, $d_3$ are listed in Table 1. 

Using the Millennium Galaxy Catalogue data
\citep[MGC;][]{mgc01,mgc05,allen06} we now derive the total $B$-band
GLF incorporating the impact of dust attenuation. 
This involves
seperating all galaxies into their bulges and disc components,
correcting their fluxes according to the formulae above, and then
rederiving their combined fluxes (note: pure elliptical systems are
not modified). Fig.~2 shows the raw observed GLF (red dotted line),
the inclination corrected GLF (green dashed line), and the fully
dust-corrected GLF (blue solid line, i.e., corrected for both the
empirically verified inclination-dependent attenuation plus the
residual face-on attenuation determined by the model). The
corresponding Schechter function fits to the GLFs, obtained using a
standard step-wise maximum likelihood estimator \citep{efstathiou88},
are also shown on Fig.~2 and tabulated in Table 2. The changes due to
the dust correction are significant. Going from the original, observed
$B$-band GLF to the final, fully dust-corrected GLF, we find a
$19$$\sigma$ shift in the characteristic luminosity ($M^*$), and a
$5$$\sigma$ change in the faint-end slope (which determines the space
density of dwarf galaxies). Nearby galaxies therefore produce far more
photons in the $400$--$450$~nm range ($B$-band) than previously
supposed. In fact, after first removing the contribution to the GLF
due to ellipticals ($0.14$ $h_{70}$~L$_{\odot}$~Mpc$^{-3}$;
\citealp{mgc05}), only $(58\pm5)$\% of $B$-band photons escape from
the nearby spiral galaxy population into the IGM (or $60\pm5$\% if one
includes the ellipticals).

\subsection{Extrapolating the impact of dust on the GLF to other bandpasses}
The above result provides us with a single dust corrected datum for
the CSED. Unfortunately we cannot repeat this measurement at other
wavelengths because we lack the appropriate imaging data at this time.
We can however resort to a simplification: For all possible B/T values
from 0 to 0.8 we derive, using the equations given above, the implied
photon escape fraction integrated over $\cos(i)$. We then adopt as the
effective mean B/T, that model galaxy whose photon escape fraction is
the same as that determined from our GLF analysis. This yields an
intrinsic $\langle B/T \rangle = 0.13^{+0.22}_{-0.13}$. Essentially,
this is an effective average $B/T$ value that corresponds to the
volume and luminosity weighted average of the individual escape
fractions of all the galaxies in our sample in the absence of dust.
%Since the luminosity density is dominated by galaxies near $M*$,
%the above observed $\langle B/T \rangle$ ratio is not surprisingly
%very close to the average B/T of $M*$ galaxies.

The above has provided us with an effective average galaxy that
represents the galaxy population at large in the $B$-band. To now
derive the equivalent effective average galaxy at other wavelengths we
need to know the mean bulge and disc colours in order to modify the
bulge-to-total ratio accordingly. For example, galaxy bulges are
typically red (relative to the disc) so as we move towards longer
wavelengths we expect the canonical $\langle B/T \rangle$ ratio to
rise a little. To obtain the mean bulge and disc colours we supplement
our $B$-band MGC data with multi-wavelength data provided by the
overlapping Sloan Digital Sky Survey \citep{mgc05,mgc07};
\citealp{mgc01}; \citealp{allen06}). Using the median colours for
bulge-only or disc-only systems we transpose our canonical
bulge-to-total ratio to the SDSS bandpasses: $\langle B/T \rangle =
0.11$ ($u$), $0.14$ ($g$), $0.14$ ($r$), $0.14$ ($i$), $0.16$
($z$). For $JHK$ we adopt the $z$-band $\langle B/T \rangle$ ratio and
for the UV range we adopt $\langle B/T \rangle=0$ (which implicitly
assumes that star-formation has ceased in the bulge regions). 
%Note that the $\langle B/T \rangle$ changes are relatively modest.

\subsection{The photon escape fraction} We can now derive the mean photon
escape fraction at any wavelength using the bulge and disc
attenuation-inclination relations predicted by our calibrated dust
model (see Section 2) coupled with the above $\langle B/T \rangle$
values. Fig.~3 displays the corresponding photon escape fractions for
a variety of bandpasses averaged over all viewing angles. To
understand how critically this depends on the adopted $\langle B/T
\rangle$ values we also show on Fig.~3 the photon escape fractions
that result from assuming, in each bandpass, the extreme values of
$\langle B/T \rangle = 0$ (pure disc, upper dotted line) and $\langle
B/T \rangle = 0.35$ (early/mid-type disc galaxy, lower dotted line),
respectively. Evidently, the photon escape fractions do not depend
critically on the assumed "canonical" values of $\langle B/T \rangle$
and hence our shortcut method should be considered robust (i.e.,
Fig.~3 is essentially a direct prediction from the MGC calibrated dust
model with little dependence on $\langle B/T \rangle$).

\subsection{The Cosmic Energy Spectrum} 
The values shown in Fig.~3 can be used to derive the CSED corrected
for dust attenuation. A compendium of recent GLF measurements
(\citealp{mgc07}; \citealp{budavari05}; \citealp{blanton03};
\citealp{kochaneck01}; \citealp{bell03}; \citealp{babbedge06};
\citealp{huang07}; \citealp{takeuchi06}) based on nearby samples (all
allegedly complete, corrected to redshift zero, and converted to $H_0
=70$~km~s$^{-1}$~Mpc$^{-1}$, but uncorrected for dust attenuation) are
shown in Fig.~4. Where necessary these have been converted from
luminosity density units to energy density units and together they
span the wavelength range from the far-UV to the far-IR. Also shown on
Fig.~4 is our single fully dust-corrected $B$-band luminosity density
value (from Table 2), which lies significantly above the previous
uncorrected estimates.

Using the photon escape fractions from Fig.~3 we now correct the
attenuated GLF measurements to produce, for the first time, the
unattenuated CSED (grey data points on Fig.~4). This constitutes the
actual spectral energy output in the Universe today due to the total
integrated starlight before dust attenuation. To calculate the total
energy of starlight versus that which escapes into the IGM we need to
integrate over these two datasets. In order to interpolate across the
full far-UV, optical, and IR wavelength range we adopt a recent
stellar synthesis model (PEGACE, \cite{pegace} see \cite{baldry03} for
details of the modelling) which provides a reasonable fit to both the
attenuated (orange line) and, when corrected, the unattenuated (black
line) CSED.  Integrating these two curves yields a total stellar
energy output of $(1.6\pm0.2) \times 10^{35}$~W~Mpc$^{-3}$ of which
$(0.9 \pm 0.1) \times 10^{35}$~W~Mpc$^{-3}$ escapes into the IGM. Note
that these numbers take into account a $10$\% component of the CSED
longwards of $400$~nm due to ellipticals \citep{mgc05}. For the CSED
to be in equilibrium (and energy conserving) the difference between
these two energy values, $(0.7\pm0.2) \times 10^{35}$~W~Mpc$^{-3}$,
should not exceed the total emission from dust in the far-IR, the
traditional sticking point for optically thin models.  Fig.~4 also
shows a model of the dust emission \citep[red line;][]{dale02} that
reproduces fairly well the observed data
%\footnote{Note that both the
%PEGACE model to the optical data and the far-IR dust model are simply
%being used to provide a continuous description of the energy output
%for the purpose of integrating across the available data. The physical
%motivation and accuracy of these models are not particularly relevant,
%(they simply provide good functional fits).}  
Integrating the far-IR
curve we obtain a total radiant dust energy of $(0.6 \pm 0.1) \times
10^{35}$~W~Mpc$^{-3}$, in excellent agreement with our prediction.
This provides independent support that the high value for the central
face-on opacity is correct, and implies that significant corrections
to the observed flux of individual galaxies are necessary. This result
also leaves little or no room for other sources of dust heating in the
nearby Universe, such as active galactic nuclei, and provides the
first fully reconciled estimate of the CSED.

\section{Discussions and Ramifications} 
This work has reconciled three apparently inconsistent observations,
namely; the severe attenuation-inclination relation seen in the MGC
data \citep{mgc07}; the conclusion that inter-arm regions are
optically thin \citep{white00,holwerda07}; and the relatively high
far-IR dust emission. The TP-model achieves this by incorporating
distinct dust components that allow for optically thin inter-arm and
outer regions coupled with an optically thick central region. This
relatively simple development,
%and perhaps obvious given the high
%opacity of the Galactic centre, 
has a number of far-reaching
ramifications. Firstly, all basic measurements of galaxy fluxes that
do not correct for dust attenuation will require significant revision
(i.e., $0.2$--$2.5$~mag in $B$) depending on an individual galaxy's
inclination, bulge-to-total ratio and wavelength of
observation. Second, many galaxies will contain heavily embedded
bulges due to the centrally concentrated dust in their discs. This can
easily lead to significant errors in optical estimates of their fluxes
and stellar masses; this is because although stellar mass estimates do
correct for optically thin dust attenuation they cannot, of course,
correct for mass hidden behind an entirely optically thick
screen. Finally, dust attenuation could conceivably play a part in the
morphology-density relation and the proposed transformation of disc
galaxies from late- to early-type as they enter the cluster
environment. For example, using the corrections provided, it is easy
to show that an Sab galaxy at the median inclination ($60^\circ$) will
see its observed $B/T$ change from $0.3$ to $0.6$, its color get
significantly redder, and its luminosity get brighter, if all its dust
was to be removed. 
%Dust attenuation is clearly a major issue which
%demands our attention if we are to robustly compare galaxy samples
%across different environments, wavelength ranges, orientations, and
%epochs.
%The corrections given provide the starting point to this
%process.

\acknowledgments Richard Tuffs is grateful for the support of a
Livesey Award whilst working on this paper at UCLan.  
%The Millennium
%Galaxy Catalogue consists of imaging data from the Isaac Newton
%Telescope and spectroscopic data from the Anglo Australian Telescope,
%the ANU 2.3m, the ESO New Technology Telescope, the Telescopio
%Nazionale Galileo, and the Gemini Telescope. The survey has been
%supported through grants from the Particle Physics and Astronomy
%Research Council (UK) and the Australian Research Council (AUS). The
%data and data products are publicly available from
%http://www.eso.org/$\sim$jliske/mgc/ or on request from J.~Liske or
%S.P.~Driver.

{}

\clearpage

\begin{table}
\caption{Coefficients for use in
Equations 1 and 2 to provide bulge and disc attenuation corrections
(with $<5$\% error in the predicted values due to the adopted fitting
function).}

\begin{tabular}{ccccccc} \hline
Bandpass & $b_1$ & $b_2$ & $b_3$ & $d_1$ & $d_2$ & $d_3$ \\ \hline \hline
$u$ & $1.10$ & $0.95$ & $2.18$ & $0.45$ & $2.31$ & $3.42$ \\
$B$ & $0.89$ & $1.27$ & $1.73$ & $0.24$ & $1.20$ & $2.73$ \\
$g$ & $0.83$ & $1.29$ & $1.71$ & $0.22$ & $1.18$ & $2.74$ \\
$r$ & $0.63$ & $1.33$ & $1.73$ & $0.16$ & $1.10$ & $2.80$ \\
$i$ & $0.48$ & $1.35$ & $1.84$ & $0.11$ & $1.03$ & $2.89$ \\
$z$ & $0.38$ & $1.35$ & $1.84$ & $0.09$ & $0.96$ & $2.98$ \\
$J$ & $0.25$ & $1.22$ & $2.26$ & $0.06$ & $0.80$ & $3.21$ \\
$H$ & $0.18$ & $1.02$ & $2.43$ & $0.05$ & $0.64$ & $3.51$ \\
$K$ & $0.11$ & $0.79$ & $2.77$ & $0.04$ & $0.46$ & $4.23$ \\ \hline
\end{tabular}
\end{table}

\begin{table*}
\footnotesize
\caption{Derived Schechter luminosity function parameters for the MGC
with varying degrees of dust attenuation corrections.}

\begin{tabular}{lcccc} \hline
\footnotesize
Sample & $M^* - 5 \log h_{70}$ &  $\alpha$ & $\phi_*$ &  $j_B$ \\
      & [mag] &
     & [$10^{-3}$ $h^3_{70}$ Mpc$^{-3}$ ($0.5$~mag)$^{-1}$]
           & [$10^8$ $h_{70}$ $L_{\odot}$ Mpc$^{-3}$]
      \\ \hline \hline
No dust corr. & $-20.57\pm0.04$ & $-1.14\pm0.03$ & $6.7\pm0.3$ & $1.9\pm0.2$ \\
Inclination corr. & $-20.78\pm0.04$ & $-1.16\pm0.03$ & $7.0\pm0.3$ & $2.4\pm0.2$ \\
Incl.+face-on corr. & $-21.32\pm0.05$ & $-1.32\pm0.02$ & $4.8\pm0.3$ & $3.1\pm0.5$ \\ \hline
\end{tabular}
\end{table*}

\clearpage

\begin{figure}
\vspace{-3.5cm}
\plotone{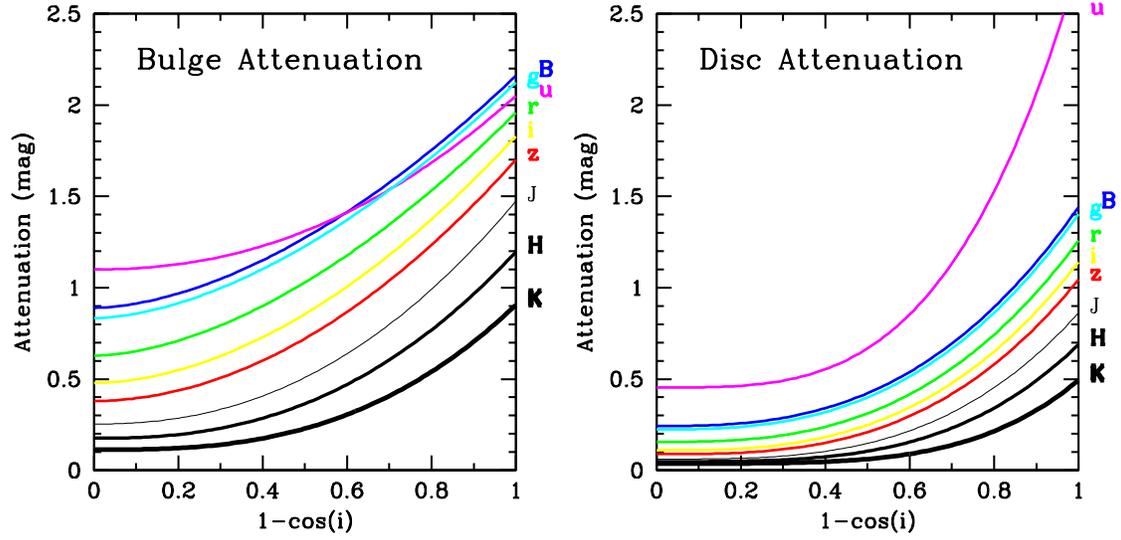}
\caption{The dust attenuation--inclination relations for galaxy discs
(right) and bulges (left) shown for a variety of bands (as indicated
on the right hand sides) as predicted by the TP-model calibrated on
$B$-band disc data \citep{mgc07}. The non-zero attenuation at $1 -
\cos(i) = 0$ is the residual face-on attenuation predicted by the
model.}
\end{figure}

\clearpage

\begin{figure}
\vspace{-2.5cm}
\plotone{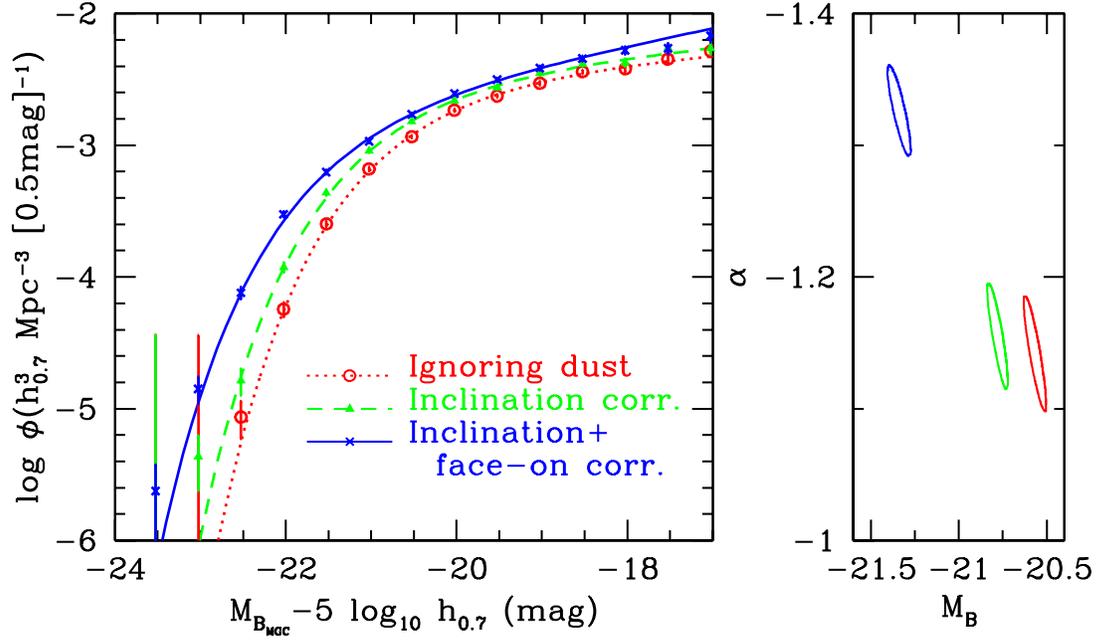}
\caption{Main panel: the $B$-band galaxy luminosity function; ignoring
all consideration of dust (dotted line), after consideration of the
empirically derived attenuation--inclination relation only (dashed
line), and after a full treatment of dust attenuation (including the
face-on model correction; solid line). Side panel: the projected
$3\sigma$ error contours for two of the three fitting parameters,
$\alpha$ (the faint-end slope) and $M^*$ (the characteristic
luminosity).}
\end{figure}

\clearpage

\begin{figure}
\plotone{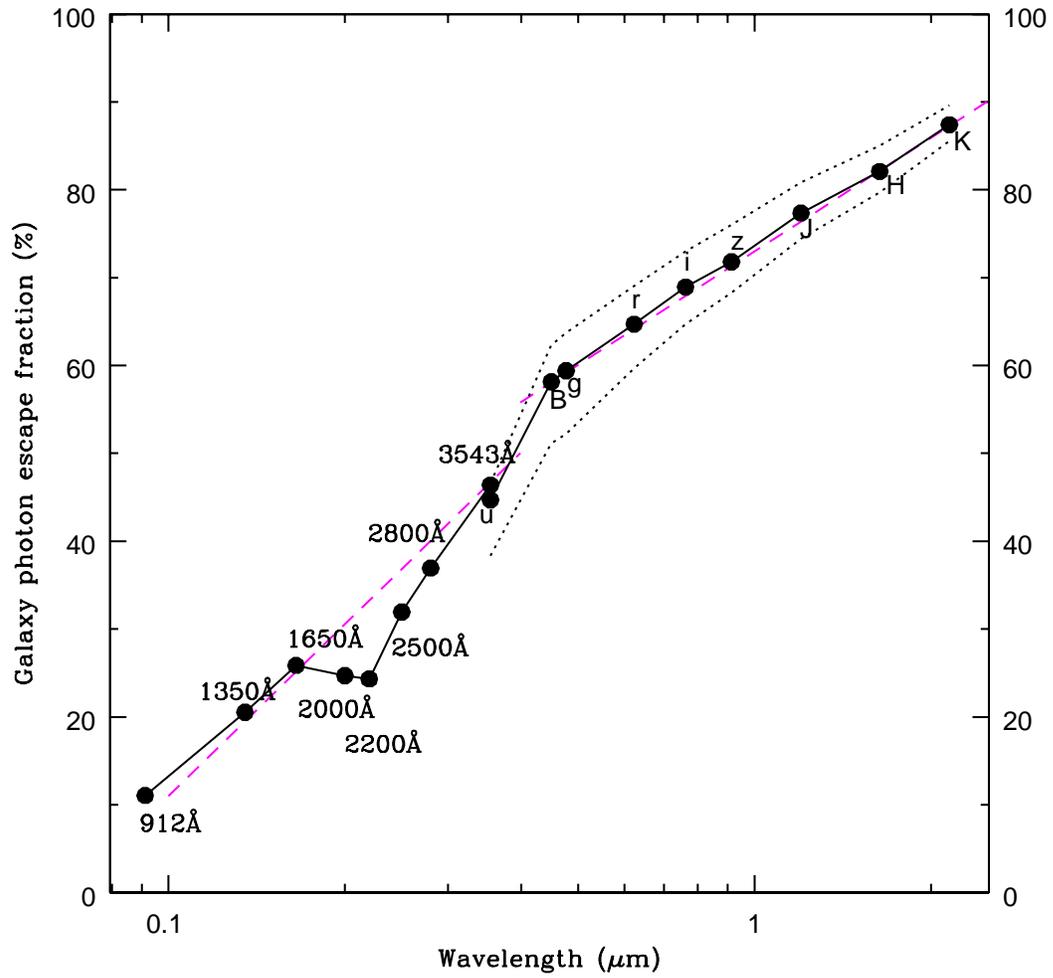}
\caption{The global photon escape fraction averaged over all
  inclinations versus wavelength. The dotted lines show estimated error
  boundaries (see text).}
\end{figure}

\clearpage

\begin{figure}
\plotone{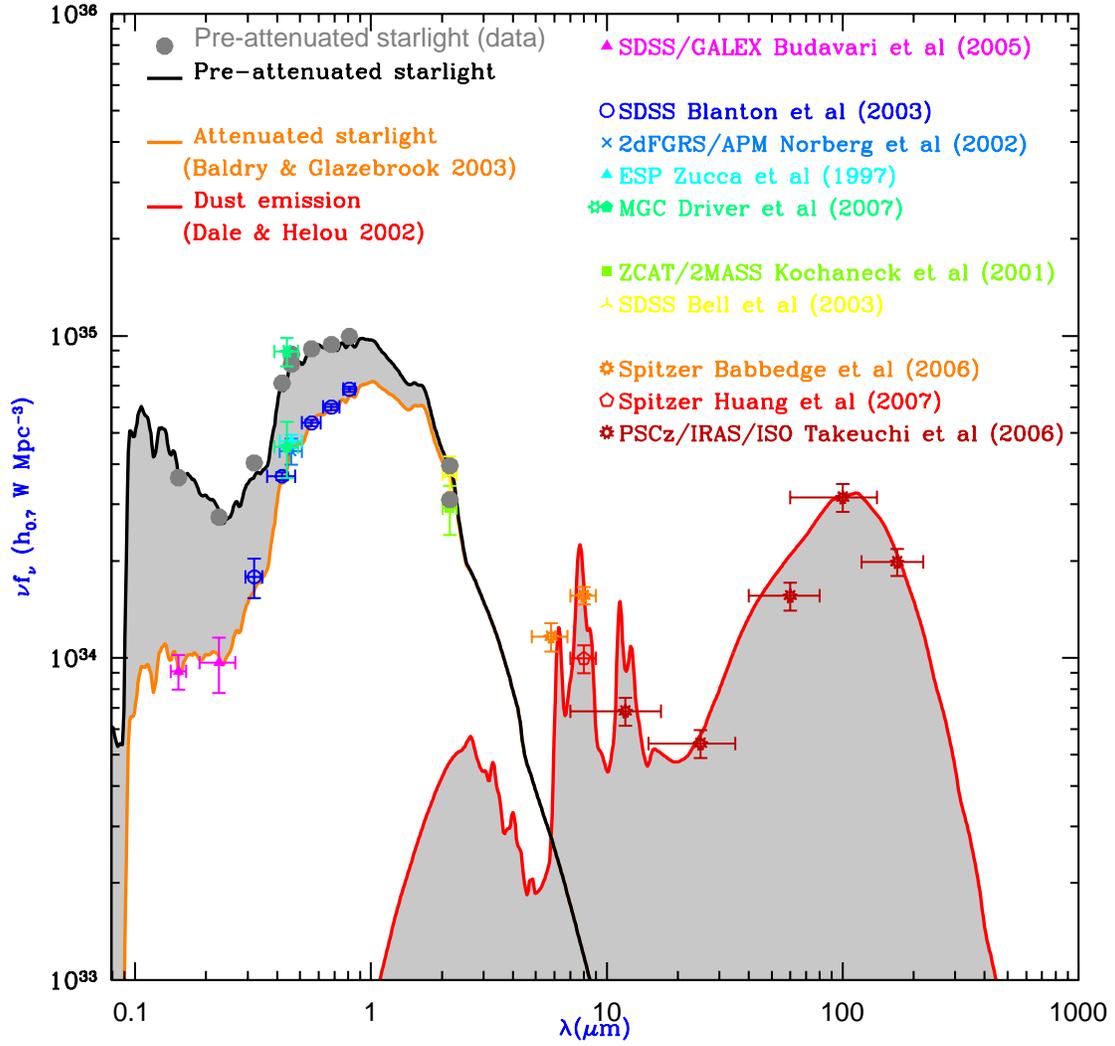}
\caption{The cosmic energy output covering the region dominated by
starlight (left peak) and by dust emission (right peak). The orange
line shows the observed (uncorrected) cosmic energy output from the
total nearby galaxy population, while the black line shows the same
after correction for the fraction of photons attenuated by dust. The
discrepancy in the integrals over these two curves yields the total
energy of starlight lost to heating of the dust grains. If starlight
is the only source of dust heating then this energy loss must equal
the total radiant energy of the dust emission (i.e., the two shaded
regions must and do contain equal energy).}
\end{figure}

%\begin{figure}
%\plotone{fig1.eps}
%\includegraphics[width=\columnwidth]{fig1.eps}
%\caption{A schematic of our dust model indicating the bulge (red) and
%disc (orange) stellar distributions along with the three dust
%components which consist of: an optically thin disc (black); an
%optically thick disc (shaded); and a clumpy component (stars). For
%completeness the equations we use to model these components are also
%shown.}
%\end{figure}


\begin{thebibliography}{}
\bibitem[Allen et al.(2006)]{allen06} Allen, P., Driver, S.P., Graham,
  A.W., Cameron, E., Liske, J., Cross, N.J.G., De Propris, R.\ 2006, 
  MNRAS, 371, 2
\bibitem[Babbedge et al.(2006)]{babbedge06} Babbedge, T.S.R., et al.\
  2006, MNRAS, 370, 1159
\bibitem[Baldry \& Glazebrook(2003)]{baldry03} Baldry, I., Glazebrook,
  K.\ 2003, ApJ, 593, 258
\bibitem[Bell et al.(2003)]{bell03} Bell, E., McIntosh, D., Katz, N., 
  Weinberg, M.D.\ 2003, ApJSS, 149, 289
\bibitem[Bianchi et al.(2000)]{bianchi00} Bianchi, S., Davies, J.I.,
  Alton, P.B.\ 2000, A\&A, 359, 65
\bibitem[Blanton et al.(2003)]{blanton03} Blanton, M., et al.\ 2003,
  ApJ, 592, 819
\bibitem[Budavari et al.(2005)]{budavari05} Budavari, T., et al.\
 2005, ApJ, 619, 31
\bibitem[Burstein et al.(1991)]{burstein91} Burstein, D., Haynes,
 M.P., Faber, M.\ 1991, Nature, 353, 515
\bibitem[Calzetti(2001)]{calzetti01} Calzetti D.\ 2001, PASA, 113, 162
\bibitem[Cardelli et al.(1989)]{cardelli89} Cardelli, J.A., Clayton,
 G.C., Mathis, J.S.\ 1989, ApJ, 345, 245
\bibitem[Choi et al.(2007)]{choi07} Choi Y., Park C., Vogeley M.S., 2007, ApJ, 658, 884
\bibitem[Dale \& Helou(2002)]{dale02} Dale, D.A., Helou, G.\ 2002,
 ApJ, 576, 159
\bibitem[Disney et al.(1989)]{disney89} Disney, M.J., Davies, J.I.,
  Phillipps, S.\ 1989, MNRAS, 239, 939
\bibitem[Driver et al.(2005)]{mgc05} Driver, S.P., Liske, J., Cross,
  N.J.G., De Propris, R., Allen, P.D.\ 2005, MNRAS, 360, 81
\bibitem[Driver et al.(2007)]{mgc07} Driver, S.P., Popescu, C., Tuffs,
  R.J., Graham, A.W., Liske, J., Allen, P.D., De Propris R.\ 2007,
  MNRAS, 379, 1022
\bibitem[Efstathiou et al.(1988)]{efstathiou88} Efstathiou, G., Ellis,
  R.S., Peterson, B.A.\ 1988, MNRAS, 232, 431
\bibitem[Felten(1977)]{felten77} Felten, J.E.\ 1977 AJ, 82, 861
\bibitem[Fioc \&Rocca-Volmeragen(1997)]{pegace} Fioc, M.,
Rocca-Volmerange, B., 1997, A\&A, 326, 950
\bibitem[Giovanelli et al.(1995)]{giovanelli95} Giovanelli, R., et
  al.\ 1995, AJ. 110, 1059
\bibitem[Holwerda et al.(2007)]{holwerda07} Holwerda, B.W., Keel,
  W.C., Bolton, A.\ 2007, AJ, 134, 2385
\bibitem[Hopkins \& Beacom(2006)]{hopkins06} Hopkins, A.M., Beacom,
  J.F.\ 2006, ApJ, 651, 142
\bibitem[Huang et al.(2007)]{huang07} Huang, J.-S., et al.\ 2007, 
  ApJ, 664, 840
\bibitem[Kochanek et al.(2001)]{kochaneck01} Kochanek, C.S.\ 2001, 
  ApJ, 560, 566
\bibitem[Liske et al.(2003)]{mgc01} Liske, J., Lemon, D.J., Driver,
  S.P., Cross, N.J.G., Couch, W.J.\ 2003, MNRAS, 344, 307
\bibitem[Maller et al.(2008)]{maller08} Maller A., Berlind A.A., Blanton M.R., Hogg D.W., 2008, ApJ, submitted (astro-ph/0801.3286) 
\bibitem[Misiriotis et al.(2001)]{misiriotis01} Misiriotis, A.,
  Popescu, C.C., Tuffs, R.J., Kylafis, N.D.\ 2001, A\&A, 372, 775
\bibitem[M\"ollenhoff et al.(2006)]{mollenhoff06} M\"ollenhoff, C.,
  Popescu, C.C., Tuffs, R.J.\ 2006, A\&A, 456, 941
\bibitem[Padilla \& Strauss(2008)]{padilla08} Padialla N.D., Strauss M.A., 2008, MNRAS, submitted (astro-ph/0802.0877)
\bibitem[Popescu et al.(2000)]{popescu00} Popescu, C.C., Misiriotis,
  A., Kylafis, N.D., Tuffs, R.J., Fischera, J.\ 2000 A\&A, 362, 138
\bibitem[Popescu \& Tuffs(2007)]{popescu07} Popescu, C.C., Tuffs, R.J.\
  2007, arXiv:0709.2310v1
\bibitem[Primack et al.(2005)]{primack05} Primack, J., Bullock, J.S.,
  Somerville, R.S.\ 2005, {\sl Observational Gamma-ray Cosmology}, in
  Proc of AIP, Vol 745, Eds: F.A.~Aharonian, H.J.~V\"olk, \& D.~Horns (Publ: AIPC), p22-33
\bibitem[Schechter(1976)]{schechter76} Schechter, P.\ 1976, ApJ, 203, 297
\bibitem[Seares(1931)]{seares31} Seares, F.H.\ 1931, PASP, 43, 371
\bibitem[Shao et al.(2007)]{shao07} Shao Z., Xiao W., Shen S., Mo H.J., Xia X., Deng Z., 2007, ApJ, 659, 1159
\bibitem[Silva et al.(1998)]{silva98} Silva, L., Granato, G.L.,
  Bressan, A., Danese, L.\ 1998, ApJ, 509, 103
\bibitem[Takeuchi et al.(2006)]{takeuchi06} Takeuchi, T.T., Ishii,
 T.T., Dole, H., Dennefeld, M., Lagache, G., Puget, J.-L.\ 2006, A\&A, 448, 525
\bibitem[Tuffs et al.(2004)]{tuffs04} Tuffs, R.J., Popescu, C.C.,
  V\"olk, H.J., Kylafis, N.D., Dopita M.A.\ 2004, A\&A, 419, 821
\bibitem[Unterborm \& Ryden(2008)]{unterborn08} Unterborn C.T., Ryden B.S., 2008, ApJ, submitted (astro-ph/0801.2400) 
\bibitem[Valentijn(1990)]{valentijn90} Valentijn, E.A.\ 1990, Nature, 346, 153
\bibitem[White et al.(2000)]{white00} White (III), R.E., Keel, W.C.,
  Conselice, C.J.\ 2000, ApJ, 542, 761
\bibitem[Xilouris et al.(1999)]{xilouris99} Xilouris, E.M., Byun,
  Y.I., Kylafis, N.D., Paleologou, E.V., Papamastorakis, J.\ 1999,
  A\&A, 344, 868

\end{thebibliography}
\end{document}